\documentclass[a4paper]{article}
\usepackage{cancel}
\usepackage{tikz}

\usepackage[english]{babel}
\usepackage[utf8x]{inputenc}
\usepackage[T1]{fontenc}

\usepackage[a4paper,top=3cm,bottom=2cm,left=3cm,right=3cm,marginparwidth=1.75cm]{geometry}

\usepackage{amsmath}
\usepackage{graphicx}
\usepackage[colorinlistoftodos]{todonotes}
\usepackage[colorlinks=true, allcolors=blue]{hyperref}

\title{Solving the Buyer and Seller's Dilemma: A Dual-Deposit Escrow Smart Contract for Provably Cheat-Proof Delivery and Payment for a Digital Good without a Trusted Mediator}

\author{Aditya Asgaonkar\footnote{Visiting student from BITS-Pilani, Goa, India; can also be reached at f20150043@goa.bits-pilani.ac.in }, Bhaskar Krishnamachari\\
 Center for Cyber-Physical Systems and the Internet of Things\\
 Viterbi School of Engineering, University of Southern California\\
 \{aa\_192, bkrishna\}@usc.edu}

\date{\today}

\begin{document}
\maketitle

\begin{abstract}
A fundamental problem for electronic commerce is the buying and selling of digital goods between individuals that may not know or trust each other. Traditionally, this problem has been addressed by the use of trusted third-parties such as credit-card companies, mediated escrows, legal adjudication, or reputation systems. Despite the rise of blockchain protocols as a way to send payments without trusted third parties, the important problem of exchanging a digital good for payment without trusted third parties has been paid much less attention. We refer to this problem as the Buyer and Seller's Dilemma and present for it a dual-deposit escrow trade protocol which uses double-sided payment deposits in conjunction with simple cryptographic primitives, and that can be implemented using a blockchain-based smart contract. We analyze our protocol as an extensive-form game and prove that the Sub-game Perfect Nash Equilibrium for this game is for both the buyer and seller to cooperate and behave honestly. We address this problem under the assumption that the digital good being traded is known and verifiable, with a fixed price known to both parties. %In case of unverifiable goods and asymmetric information between parties, the problem remains open.     
\end{abstract}

\section{Introduction}

A fundamental problem for electronic commerce has been the exchange of a digital good for payment. The earliest solutions for this problem date back at least to the earliest days of the world wide web~\cite{timmers1999electronic}, online stores accepting credit card payments for downloadable content. Because the exchange of the good and payment cannot happen simultaneously, there is an inherent tension and need for trust in the trade --- the seller must trust that the buyer will pay and the buyer must trust that the seller must deliver. Traditionally, this need for trust has been addressed by introducing a trusted third party --- this could be a credit card company, a third party mediator for an escrow~\cite{Goldfeder2017EscrowPF}, legal adjudication or arbitration of disputes~\cite{ponte2005cyberjustice} or the use of a reputation system to build trust by allowing parties to gain some understanding of the prior behavior of the other~\cite{resnick2002trust}. Indeed underground economies where such trust is difficult to obtain are often rife with scams~\cite{herley2010nobody}. 

The blockchain revolution that was ushered in by the publication of the technical paper on Bitcoin~\cite{nakamoto2008bitcoin} has allowed for the first time for digital payments to be made between parties without requiring a trusted third party. However, the problem of exchanging a digital good for payment without requiring a trusted third party has not been widely addressed. In principle, agreements between parties can be codified in the form of smart contracts; however, even recent work on applying smart contract to such transactions has continued to rely on third party mediation. For example, Goldfelder \emph{et al.}~\cite{Goldfeder2017EscrowPF} consider the same problem and propose the use of smart contracts that involves a third-party mediator or a group of third party mediators, while proving certain security and privacy enhancements over traditional approaches.

In contrast to that prior work, we propose here a smart contract that is deployed by the seller. This smart contract requires both the seller and the buyer to place a sufficiently high deposit into the smart contract. In our proposed protocol, seller first submits its deposit, then the buyer sends payment as well as its own deposit; the seller then sends a key to unlock the digital good. The buyer verifies the good is received and if all is well, sends an approval message to the smart contract. The deposits made by both parties are returned to them only after a successful trade is completed (the seller sends the correct good and the buyer verifies and approves it). In all other cases at least one of the parties will lose its deposit. The protocol involves no third parties at all. We analyze this dual-deposit escrow trade protocol as an extensive form game and show that honest behavior by both parties is the only subgame perfect Nash equilibrium. 

While the use of one-way security deposits to provide trust for one party with respect to the other is quite common and dates back a long time, particularly in the context of home rentals~\cite{wilson1934lease}, dual-deposits such as the scheme proposed in our protocol are not common~\footnote{A fascinating historical example of dual-deposit escrow, though, is mentioned in Julius Caesar's autobiographical \emph{Gallic Wars} (Book 2) which describes mutual hostage exchange between certain tribes as a form of diplomacy: `` all the Belgae... were entering into a confederacy against the Roman people, and giving hostages to one another.''}. However, after we wrote our first draft of this paper, we became aware of a double-deposit escrow mechanism provided by a decentralized online marketplace called BitBay~\cite{Bitbay}, which we further learned is  essentially the same as another system called BitHalo~\cite{zimbeck2015two} (both developed by  David Zimbeck, from what we were able to determine from online sources). Bigi \emph{et al.}~\cite{bigi2015validation} have formalized Bithalo into a particular double-deposit scheme they refer to as DCSP (for decentralized smart-contract protocol) and analyze it game theoretically. There are some differences from our scheme in this paper: in the BitHalo/BitBay double-deposit escrow scheme buyers and sellers both make a deposit through the client or smart-contract, then exchange the good and payment off-chain, and both deposits are released only if both parties confirm that the transaction was successful, with no other form of dispute resolution. Their scheme does not allow for the delivered digital good to be independently verified. In contrast, in our scheme the smart contract is capable of autonomous verification based on a known hash of the digital good - this allows our proposed smart contract to selectively return the buyer's deposit if a complaint by a buyer is valid and to selectively return the seller's deposit if a complaint by a buyer is found to be invalid. 

A closely related but fundamentally different problem that has been addressed previously is that of \emph{Atomic Swaps}~\cite{tiernolan} used in blockchain systems to exchange on-chain assets or tokens between users using smart-contracts. This can be accomplished in a decentralized manner more easily because the movement of such assets corresponds merely to a change of state (such as changing the owner of assets) on the blockchain. In contrast, sending a digital good to the buyer in an encrypted manner in exchange for payment so that only the buyer can decrypt it does not represent merely a change of state on the blockchain.

% Solutions for atomic swap rely on revealing a secret on the chain and the corresponding state change is transparent to all parties; in contrast in our problem of sending a digital good it is necessary for the good to be kept encrypted so that no other parties can access it. Sending a digital good to the buyer, thus, does not represent merely a change of state on the blockchain. 

% In particular the asset is temporarily locked into a smart contract and released when a required condition is met. In doing so, there is no requirement to encrypt any asset to hide it from the other party until the payment is made. In contrast, when a digital good is being sent from a seller to a buyer using a smart contract, there is no way to send it to or lock it up temporarily in a smart-contract without encrypting it because otherwise the buyer may access it without paying for it by merely reading the corresponding bits. This subtle difference means that solutions for atomic swap cannot be used to address the problem of selling and buying a general digital good. 

%As the Ethereum Foundation's Virgil Griffith puts it - Ethereum is a framework that allows actors to play positive sum games that were not possible without it. There are several game theoretic problems that need to be analyzed in the light of these new blockchain technologies. In this work, we attempt to address such a game: The Buyer and Seller's Dilemma. 

\subsection{The Buyer and Seller's Dilemma}

Consider a scenario where a Seller is attempting to make a sale of a Product to the Buyer. Two transactions are bound to happen:
\begin{description}
\item[Delivery of Product]:  The Seller delivers the Product to the Buyer.
\item[Payment for Product]: The Buyer makes a payment to the Seller.
\end{description}

In any trading platform, one of these two transactions must occur first. Depending on which transaction occurs first, one of these factors of trust is introduced between the Buyer and Seller:
\begin{description}
\item[Trust of Delivery]: Trust in the Seller that if payment is made first, the Product will be delivered.
\item[Trust of Payment]: Trust in the Buyer that if the Product is delivered first, then payment will be made.
\end{description}

In nearly all existing systems, the latter transaction is not guaranteed to occur, but is incentivized through a separate entity --- an escrow account, a legal document, or a reputation system. Each of these systems have a third-party trusted actor:
\begin{description}
\item[Escrow Account]: A third-party trusted mediator is introduced who will hold the payment from the Buyer until the Seller delivers the Product.
\item[Legal Document]: A legally binding document is introduced that will make the cheating party liable to facing penalty from a judiciary system. There is inherent trust in the judiciary system.
\item[Reputation System]: A reputation system lists the Seller, and any complaints against it. A potential buyer can look up the Seller's reputation, and make the payment first. Further malicious behavior from the Seller is disincentivized through the risk of damage to reputation. There is inherent trust in the reputation system.
\end{description}

By Buyer and Seller's Dilemma, we are referring to solving the problem of trust between the Seller and Buyer of a digital good without involving a third party. The name is coined by analogy with the well-known Prisoner's Dilemma~\cite{rapoport1965prisoner}, as the participants in this game must also think about whether to cooperate (i.e., act honestly) or defect (i.e., cheat on each other). However, in this case the game that emerges is a sequential or extensive-form game because the Seller and Buyer don't move simultaneously. 

% \subsection{Objective}
% We aim to design a system that facilitates provably cheat-proof transactions of delivery and payment between a Seller and a Buyer. We do not rely on a third-party trusted mediator, instead using a smart-contract-based blockchain framework to provide our cheat-proof guarantees. The system mitigates trust factors between the two parties by providing a strong on-chain incentive for non-malicious behavior. 

\subsection{Trust, Enforceability, and Incentive}
In order to remove trusted third parties, there must be constraints that are enforced on the behavior of two participating actors. Blockchain systems provide a programmatic way to process and regulate transactions that participants propose in an autonomous manner, referred to as smart contracts~\cite{buterin2014next}. It is important to note that in the context of implementing a protocol using a blockchain-based smart contract, only those interactions which happen through the smart contract can be verified, policed or constrained. Interactions that occur outside the smart contract --- off-chain interactions between sellers and buyers --- cannot be constrained. 
%BK: This is an assumption or goal for us?
%AA: Not intended as an assumption, simply stating general facts. Although this is assumed in the later game theoretic analysis

Therefore the only way to influence the off-chain behavior of the participants is by providing on-chain incentives to the actors to conform to good behavior. The crux of our protocol, which we refer to as the dual-deposit escrow trade protocol, lies in the design of such an incentive scheme. We further conduct game theoretic analysis of this incentive scheme, and prove that it is in the best interest of both parties to display honest, mutually helpful behavior.

\section{Solution to the Buyer and Seller's Dilemma}

\subsection{Assumptions}
\begin{itemize}
\item We assume the Product being traded as any digital or physical asset that can be secured against unauthorized use through a digital key. We assume that the Product is accessible only using the digital key $d$ (or is the digital data $d$ itself). We can now use $d$ and Product interchangeably.
\item We also assume that the Buyer knows the hashed value $h(d)$ for the Product $d$, which he/she could use to verify that a received product is correct.
%AA: I think this is a VERY strong assumption. Needs discussion.

\item The Seller and Buyer both have an asymmetric key pair, with their public keys known to each other. 
\item We assume that transaction fees associated with deploying the smart contract and sending transactions to it are negligible compared to the price of the product.  
\end{itemize}

\subsection{The Dual-Deposit Escrow Trade Protocol}

\begin{figure}
\centering
\includegraphics[width=\textwidth]{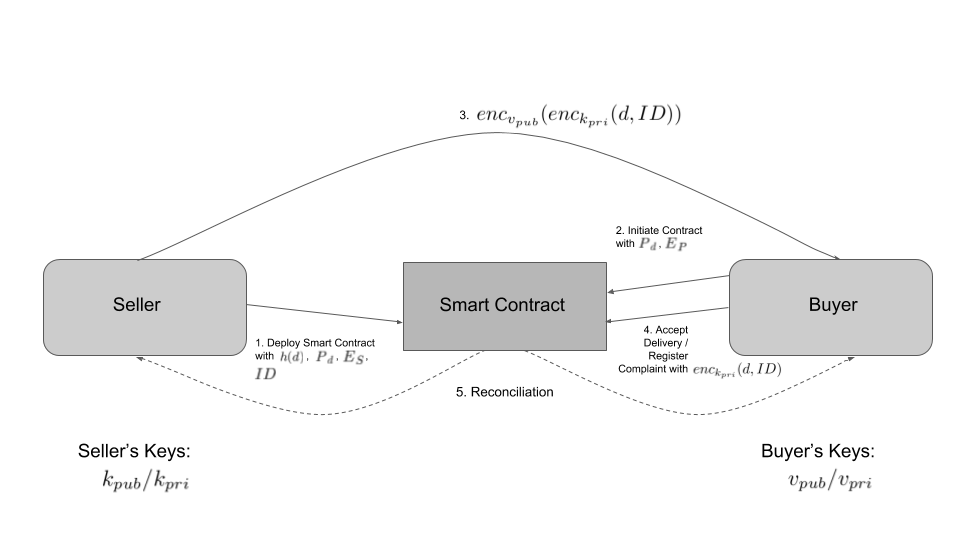}
\caption{\label{fig:arch}Proposed System Architecture}
\end{figure}
%BK: need to update figure to show deposit and payment amounts, show final resolution step

\begin{enumerate}
\item 
%The Seller publishes a smart contract that advertises its product.
%BK: the above is not essential to say here, I believe
%AA: Agreed, not essential

\textbf{Seller Deployment:} For each sale, the Seller will publish a new smart contract that includes
\begin{itemize}
\item[$P_d$]: Price of the Product
\item[$h(d)$]: Hash of the Product
\item[$ID$]: Contract Nonce
\end{itemize}
The Seller must also make a Seller Deposit $\mathcal{E}_{S}$ to the Smart Contract, which is later refunded.
%BK: let's use \mathcal{E} instead of \varepsilon which suggests a negligible deposit
%AA: Got it

%BK: the ID generated is used to prevent buyer from doing a replay attack
%AA: Yes, this is required.

\item \textbf{Buyer Initialization:}  The Buyer then initiates the Smart Contract. The Buyer must pay the price $P_d$ for the product and also make a Buyer Deposit $\mathcal{E}_{B}$, that is later refunded.

\item \textbf{Delivery:} The Seller sends the encrypted version of $d$, namely, $enc_{v_{pub}}(enc_{k_{pri}}(d, ID))$, to the interested Buyer, possibly on an off-chain channel.

\item \textbf{Accept/Reject Delivery:} The Buyer decrypts the data $d$, then hashes it to check if it matches the previously known $h(d)$. The Buyer then provides a response to the Smart Contract; in this response, it either:
\begin{itemize}
\item Accepts delivery of the Product.
\item Rejects delivery, claims that the Seller has cheated, and tries to prove it by sending $enc_{k_{pri}}(d, ID)$ to the Smart Contract.  
%BK: the ID generated by the seller in step 1 is needed. otherwise the seller is vulnerable to a replay attack by having the buyer send back some old encrypted data and claiming the seller cheated. 

% BK: There is STILL A PROBLEM, I think; what if the seller sends a random set of bits that cannot be decrypted. Or if the buyer claims the seller sent a random set of bits? 
%AA:	1) Random set of bits that cannot be decrypted: This case is especially tricky. We cannot determine who cheated if it cannot be decrypted. The Buyer may just submit garbage and claim that the Seller's data cannot be decrypted.
%		2) The Buyer claims the seller sent a random set of bits: The smart contract can decrypt the Seller's signed message, check the ID, then compare h(d) with what it knows. The Seller MUST send something that hashes to h(d), otherwise it can be determined that he cheated.

\end{itemize}
\item \textbf{Reconciliation:} This step is undertaken by the Smart Contract after hearing from the Buyer in the previous step. 
\begin{itemize} 
\item In case of Acceptance: Both the Seller Deposit and the Buyer Deposit are refunded to corresponding parties. The Seller also receives the price $P_d$ for the product.
\item In case of Complaint: The Smart Contract will decrypt $enc_{k_{pri}}(d, ID)$. If the Buyer submits a garbage string, then the Smart Contract will slash both deposits, along with the payment. Then it first compares $ID$ to ensure that the ciphertext was corresponding to this transaction. It then hashes $d$ and find $h(d)$, which is then compared with the one that the Seller uploaded while generating the smart contract.  
\begin{itemize} 
\item If a mismatch is found, then the Seller has cheated, and loses its deposit $\mathcal{E}_S$: it is used to pay for the gas consumed in this reconciliation step, and the rest is slashed (burned). The Buyer gets back its deposit as well as the payment for the product.
\item If the hashes match, then the Buyer made a frivolous complaint, and loses its deposit $\mathcal{E}_B$: it is used to pay for the gas consumed in this reconciliation step, the Product payment $P_d$ is sent to the Seller, and the rest is slashed (burned).
\end{itemize}
\end{itemize}

\end{enumerate}

\subsection{Game Theoretic Analysis}

The dynamics of this interaction between the Seller and Buyer can be modeled as an extensive form game, with the Seller playing the first move (Step 3 in the protocol description), and the Buyer playing the second move (Step 4 in the protocol description). We analyze this extensive form game to find its Subgame Perfect Nash Equilibrium, the strategy profile for both players that ensures no one has an incentive to deviate in any sub-game of the original game~\cite{fudenberg1991game}. 

We use the following labels in the game tree:
\begin{description}
\item[N, N']: Non-fraudulent (honest) behavior by the Seller and Buyer, respectively.
\item[F, F']: Falsified data submission by the Seller and Buyer, respectively. For the Seller this would correspond to sending the wrong data, but signed by it's key, in Step 3. For the Buyer it would correspond to trying to dispute the transaction with a replay attack in Step 4.
\item[G, G']: Garbage data submission by the Seller and Buyer, respectively. This corresponds to the Seller sending a string that cannot be decrypted with the corresponding public key in Step 3, or the Buyer doing so in Step 4.
\item[S]: Frivolous complaint by the Buyer. This corresponds to disputing while providing evidence of honest delivery. 
\item[R]: No response by the Buyer in Step 4
\end{description}

Also, we use $V_d$ to denote the value of the Product (from the Buyer's perspective). We are assuming that $V_d > P_d$.
%BK: this is odd, we should not need this assumption at all, it is very strong)
%AA: Alternative is that we just use P_d as the value of the data. 

%BK: I think we are assuming gas prices are negligible; we should make this explicit - in fact best not to call it gas price but rather transaction fee / smart contract deployment fee etc. to be more general than even ethereum
%AA: For the game theoretic analysis, since each possible play ends after the same number of on-chain transactions, we need not include that. We can state this in the discussion.

We first present the analysis of the payoffs for different interactions between the Buyer and the Seller:

\begin{itemize}
\item If the Buyer falsifies its response (i.e., plays $F'$) in Step 4, then regardless of the seller's actions, the Smart Contract will send the payment $P_d$ to the seller and slash the Buyer's deposit $E_B$. In the case that the Seller was honest (played $N$), then the Buyer receives the Product, and its payoff increases by $V_d$.

\item Similarly, if the Buyer submits a garbage string (plays $G'$), then the Smart Contract will slash both parties' deposit, $E_B$ and $E_S$, as well as the buyer's payment of $P_d$. In the case that the Seller was honest (played $N$), then the Buyer receives the Product, and its payoff increases by $V_d$.

\item In the case that the Buyer does not submit any acceptance or complaint in Step 4 (plays $R$), then the deposits of both parties' and the Buyer's payment of $P_d$ remain permanently locked. These can be treated as a loss in the payoffs. In the case that the Seller was honest (played $N$), then the Buyer receives the Product, and its payoff increases by $V_d$. In the case that the Seller was honest (played $N$), then the Buyer receives the Product, and its payoff increases by $V_d$.

\item If the Buyer is being honest (plays $N'$), then:
\begin{itemize}
\item If the Seller sent falsified data (plays $F$), then the Smart Contract can identify cheating on the Seller's part. The Seller's deposit is slashed and the Buyer is refunded it's deposit and payment.
\item If the Seller sent garbage to the Buyer (plays $G$), then the Buyer sends a garbage string to the Smart Contract. This results in the slashing of deposits and the payment.
\item If the Seller sent the actual data (plays $N$), then the Buyer accepts the delivery. This results in the payment to the Seller, and refund of respective deposits.
\end{itemize}

\end{itemize}

\begin{figure}[t]
\hspace*{-2.5cm}
\begin{tikzpicture}[
% 	level 1/.style={sibling distance=20em},
%   level 2/.style={sibling distance=7em},
    scale=1
]
\tikzstyle{hollow}=[circle,draw,inner sep=1.5]
\tikzstyle{solid}=[circle,draw,inner sep=1.5,fill=black]
\node[hollow, label=above:{Seller}]{}
  child[sibling distance=18em]{node[solid, label=above:{Buyer}]{}
  	[sibling distance=6em]
    child{ node{{\tiny$(P_d, -P_d-\mathcal{E}_{B})$}} 
    		edge from parent node[left] {\small F'}
    }
    child{ node{{\tiny$(-\mathcal{E}_{S}, -P_d-\mathcal{E}_{B})$}} 
    		edge from parent node[left] {\tiny G'/R}
    }
    child{ node{{\tiny$(-\mathcal{E}_{S}, 0)$}}
    		edge from parent node[right] {\small N'}
    }
    edge from parent node[above] {\small F}
  }
  child[sibling distance=18em]{node[solid, label=left:{Buyer}]{}
  	[sibling distance=6em]
    child{ node{{\tiny$(P_d, -P_d-\mathcal{E}_{B})$}}
    		edge from parent node[left] {\small F'}
    }
    child{ node{{\tiny$(-\mathcal{E}_{S}, -P_d-\mathcal{E}_{B})$}}
    		edge from parent node[right] {\tiny G'/R}
    }
    child{ node{{\tiny$(-\mathcal{E}_{S}, -P_d-\mathcal{E}_{B})$}}
    		edge from parent node[right] {\small N'}
    }
    edge from parent node[left] {\small G}
  }
  child[sibling distance=20em]{node[solid, label=above:{Buyer}]{}
  	[sibling distance=7em]
    child{ node{{\tiny$(P_d, V_d-P_d-\mathcal{E}_{B})$}}
    		edge from parent node[left] {\tiny F'/S}
    }
    child{ node{{\tiny$(-\mathcal{E}_{S}, V_d-P_d-\mathcal{E}_{B})$}}
    		edge from parent node[right] {\tiny G'/R}
    }
    child{ node{{\tiny$(P_d, V_d-P_d)$}}
    		edge from parent node[right] {\small N'}
    }
    edge from parent node[above] {\small N}
  };
\end{tikzpicture}
\caption{Game Tree for the Dual-Deposit Escrow Trading Protocol}
\label{fig:gtree}
\end{figure}
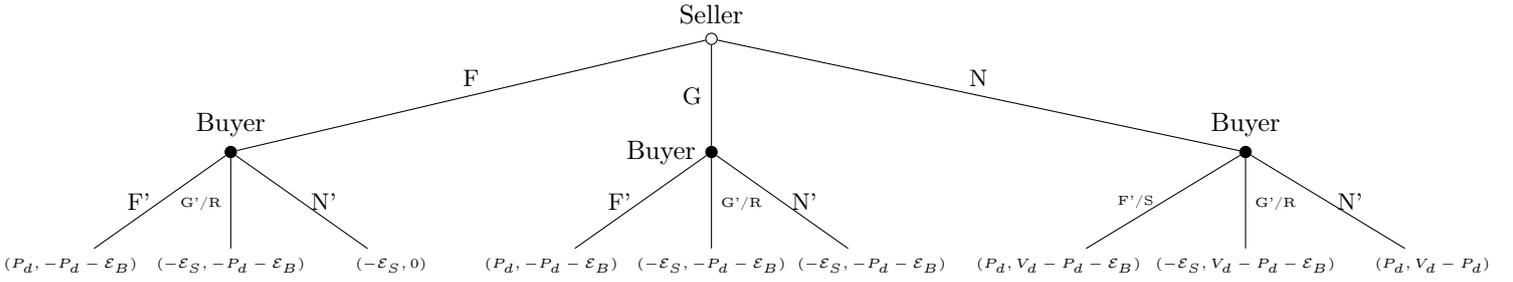

The resulting extensive form game with these payoffs is shown in Figure~\ref{fig:gtree}. It can be determined from backward induction analysis on this tree that there is only one Subgame Perfect Nash Equilibrium of $(P_d, V_d-P_d)$ that is achieved by the strategy profile (N, N'), when both parties are non-fraudulent (i.e. honest), so long as $\mathcal{E}_B, \mathcal{E}_S > 0$

% Tutorial for drawing extensive game trees
% http://www.sfu.ca/~haiyunc/notes/Game_Trees_with_TikZ.pdf

%BK: actually, something is wrong, I think, in case the buyer does something fraudulent, shouldn't the buyer's penalty be $\mathcal{E}_B - P_d - E_B$? In this case even that condition is not needed. The protocol seems to work now for any non-zero deposit (which is a bit add/concerning that it has no dependence on P_d or V_d...?)

\subsection{Safety and Liveness}

\noindent\textbf{Safety}: The presence of a unique Sub-game Perfect Nash Equilibrium with positive payoffs for both players guarantees the safety of the protocol. Further, by making the parameters $\mathcal{E}_S$ and $\mathcal{E}_B$ arbitrarily large, we can strengthen the disincentive for malicious behavior to the required standard.

\noindent\textbf{Liveness}: The players are incentivized to move forward in the trade because of the opportunity cost of the large deposit amounts locked in the smart contract. Also, we can guarantee liveness in between Step 1 and Step 2 by making a provision for the Seller to cancel the trade and refund its deposit if the buyer has not moved.

\section{Discussion and Future Work}
We presented a dual-deposit escrow trade protocol for cheat-proof transactions of payment and delivery between the two participants in the trade of a verifiable digital good. We base our cheat-proof guarantees on a game theoretic analysis of the interactions between the Seller, Buyer and Smart Contract. The safety and liveness properties of the protocol can be improved by increasing the deposit amounts. 

While our analysis is suitable for rational but selfish participants, we should note that it does not explicitly cover the case of irrationally malicious parties that are willing to take a negative payoff in order to inflict harm on the other party (although increased deposit amounts may provide some mitigation against such behavior at the cost of raising the barrier to transaction). 

We are currently in the process of implementing the proposed protocol as a smart contract on Ethereum. One challenge in such an implementation is the need for decryption in case the buyer disputes delivery; this is not a function that is expected to be used given the analysis, however it is needed for correctness of the protocol; while in theory any Turing-complete smart contract language would allow the implementation of decryption, in practice the use of an oracle might be needed for cost reasons. 

A central assumption made in this work is that the digital good being exchanged is verifiable, in particular that the buyer (or the smart contract in case of a dispute, when presented with the relevant evidence) has the ability to verify that the correct good is received --- for ease of exposition we assume this is accomplished by a hash of the digital good being known to the buyer and smart contract in advance of the transaction. The problem would be become considerably harder (if not impossible) to achieve without a trusted third party if the buyer cannot independently verify the delivery. Further, we assume that the price of the good being exchanged is already known to both parties. Problems with asymmetric information about prices fall under the broader class of ``lemon market" problems~\cite{akerlof1978market}. The implications of such lemon market problems in crypto-economic environments is worthy of further study. 

The mechanism presented here to pay for the delivery of digital goods could also in principle be used to pay for physical goods provided by the seller that are kept locked in a box that is secured by a digital key~\cite{buettgenbach2002methods}. This key could be the digital good in our description with everything else remaining the same. Once the buyer gets this key he/she can open the box to retrieve the purchased physical good (assuming that the hash of the key that is assumed to be known to both parties was generated after ensuring that the physical good is inside the locked box). 

We had earlier mentioned that solutions for atomic swap are not applicable to general payment for digital goods, however, the reverse may not be true. We are currently exploring how our approach may be potentially used for the atomic swap problem.

% $k_{pub} /  k_{pri}$ 
% $v_{pub} / v_{pri}$

%In our future work, we aim to analyze the impact of our assumption that the Buyer knows $h(d)$. 

%We also aim to identify the characteristics of digital and physical trades which can be conducted using our protocol.

\bibliographystyle{IEEEtran}
\bibliography{main}

\end{document}